\documentclass[10pt,preprint]{aastex}

\newcommand{\p}{\ensuremath{p}}
\newcommand{\nuc}{\ensuremath{\nu_{\mathrm{c}}}}

\shorttitle{GRB afterglows as probes of environment II}
\shortauthors{Starling et al.}

\begin{document}

\title
{Gamma-Ray Burst Afterglows as Probes of Environment and Blastwave Physics II:
  the Distribution of $p$ and Structure of the Circumburst Medium}

\author{R. L. C. Starling\altaffilmark{1,2,*}, A. J. van der
  Horst\altaffilmark{2}, E. Rol\altaffilmark{1},
  R. A. M. J. Wijers\altaffilmark{2}, C. Kouveliotou\altaffilmark{3},
  K. Wiersema\altaffilmark{2}, P. A. Curran\altaffilmark{2} and P. Weltevrede\altaffilmark{2}}
\altaffiltext{1}{Dept. of Physics and Astronomy, University of Leicester, University Road, Leicester LE1 7RH, UK}
\altaffiltext{2}{Astronomical Institute `Anton Pannekoek', University of Amsterdam, Kruislaan 403, 1098 SJ Amsterdam, The Netherlands}
\altaffiltext{3}{NASA Marshall Space Flight Center, NSSTC, VP-62, 320 Sparkman Drive, Huntsville, AL 35805, USA}
\altaffiltext{*}{email rlcs1@star.le.ac.uk}






\begin{abstract}

We constrain blastwave parameters and the circumburst media of a subsample of
ten {\it BeppoSAX} Gamma-Ray Bursts. 
For this sample we derive the values of the injected electron energy distribution index, \p, 
and the density structure index of the circumburst medium, $k$, 
from simultaneous spectral fits to their X-ray, optical and nIR afterglow data. 
The spectral fits have been done in count space and include the effects of metallicity, 
and are compared with the previously reported optical and X-ray temporal behavior. 
Using the blastwave model and some assumptions which include
on-axis viewing and standard jet structure, constant blastwave energy and no
evolution of the microphysical parameters, we find a mean value of $p$ for the sample as a whole of $2.04^{+0.02}_{-0.03}$. 
A statistical analysis of the distribution demonstrates that the $p$ values in this sample 
are inconsistent with a single universal value for $p$ at the 3$\sigma$ level or greater, 
which has significant implications for particle acceleration models. 
This approach provides us with a measured distribution of circumburst density structures 
rather than considering only the cases of $k=0$ (homogeneous) and $k=2$ (wind-like). 
We find five GRBs for which $k$ can be well constrained, and in four of these cases 
the circumburst medium is clearly wind-like. The fifth source has a value of $0\le k\le 1$, 
consistent with a homogeneous circumburst medium.

\end{abstract}

\keywords{gamma rays: bursts}


\section{Introduction}

Since the discovery of GRB afterglows in 1997 
\citep{costa1997:nature387:783,vanparadijs1997:nature386:686,frail1997:nature389:261}, 
their relatively longer duration and broader wavelength coverage compared with the prompt emission 
has made them the most accessible and arguably the most profitable area for observational GRB studies.
Their non-thermal spectra and lightcurves can generally be well described by the fireball model 
\citep{cavallo1978:mnras183:359,rees1992:mnras258:41,meszaros1997:apj476:232,wijers1997:mnras288} which describes a decelerating, 
highly relativistic outflow, the so-called blastwave, interacting with a surrounding external medium. 
Adaptations have been made to accommodate a non-uniform external medium \citep{meszaros1998:apj499:301,chevalier1999:apj520:29} 
and the fact that the outflow is collimated \citep{rhoads1997:apj487:487,rhoads1999:apj525,sari1999:apj519:17}. 
The evidence for collimation comes from energy considerations, and directly from observations 
of an achromatic break in afterglow lightcurves, the so-called jet-break. 
The jet-break is an indication of the time at which sideways spreading of the jet begins to become important, 
combined with the fact that the effect of relativistic beaming starts to be noticeable within the jet. 
For the long-duration GRBs, which most probably originate in the core-collapse of a massive star 
\citep{woosley1993:apj405:273,macfadyen2001:apj550,woosley2006:rmp74:1015,hjorth2003:nature423:847}, 
a $n(r) \propto r^{-2}$ density profile from a stellar wind would not be unexpected for the external medium. 
Its density can be probed through the temporal decays of the afterglow and, 
contrary to expectations given their stellar progenitors, long GRBs studied thus far do not all require 
a wind-like $n(r) \propto r^{-2}$ density profile in the local external medium 
\citep[e.g.][]{panaitescu2002:apj571}. 

The synchrotron nature of the blastwave radiation produces spectra and lightcurves comprising of a set of 
power laws with characteristic slopes and frequencies. Accurate measurements of these observed 
quantities allow the underlying parameters of the blastwave to be determined 
\citep[e.g.][]{wijers1999:apj523:177,panaitescu2001:apj560:49}. 
Optical through X-ray spectra provide the opportunity to measure the index of the input power law 
energy distribution of electrons, \p, potentially the index $k$ of the density profile of the circumburst medium, 
$n(r) \propto r^{-k}$, and in some cases also the cooling break frequency $\nu_{\mathrm c}$ --- 
the frequency of electrons whose radiative cooling time is equal to the dynamical timescale of the blastwave.

If the microphysics of all GRBs is assumed to be the same, the range of values of 
the input electron energy distribution indices should be narrow. 
However, the directly measurable data that lead to the parameter \p, such as the break frequencies 
and power-law slopes of the spectra and lightcurves, are dependent on various other factors, 
like the circumburst density, the fraction of energy in electrons ($\epsilon_e$) and magnetic fields ($\epsilon_B$), 
and simply the total amount of energy, making it more difficult to obtain a consistent value for \p. 
Further, determination of these parameters so far works most successfully for the later-time afterglow considered here: 
before approximately 0.1 days the observed lightcurves and spectra are likely some mixture of the prompt emission 
(attributed to internally colliding shocks) and afterglow (the external shock), see e.g., \citet{nousek2006:apj642:389} 
and \citet{obrien2006:apj647:1213}. 
These authors discuss the `canonical' X-ray lightcurve for {\it Swift} bursts, which has three phases, 
the final phase beginning between 0.01 and 0.12 days after trigger \citep{nousek2006:apj642:389} and showing the type
of steady decay seen in the pre-{\it Swift} era. The advantage of using
late-time data only is 
that we can reliably restrict ourselves to the slow cooling regime in modeling the blastwave, 
where the injection frequency of the electrons is well below their cooling frequency. 
The measurements we perform in this study are at late enough epochs 
\citep[see Table 1 in][hereafter Paper I]{starling2006astro.ph.10899}, 
that they should not be affected by any prompt emission components.

A potential problem is that occasionally values of \p\ below 2 have been found 
\citep[e.g. for GRB\,010222,][]{masetti2001:aa374:382,stanek2001:apj563:592}. 
This requires a cut-off at the high-energy end of the distribution of the electrons, 
and adaptations have been made for such cases \citep{bhattacharya2001:basi29:107,panaitescu2001:apj560:49}. 
The details of these adaptations, however, are still under debate, since the evolution of the high-energy cutoff 
is not well constrained. Since the lowest values for \p\ we find are $\sim 2$, 
we do not take any adaptations for this effect into account. 

For several bursts studied here the underlying parameter set has been measured independently. 
Determinations for sets of GRBs have been made by e.g. 
\citet{wijers1999:apj523:177,panaitescu2001:apj560:49,yost2003:apj597:459,gendre2005:aa430:465,granot2006:mnras370:1946}. Unfortunately, only a small fraction of GRBs have measurements in all wavebands, most notably GRB\,970508 
\citep[e.g.][Van der Horst \& Wijers in preparation]{wijers1999:apj523:177,panaitescu2002:apj571,yost2003:apj597:459} 
and GRB\,030329 \citep[e.g.][]{berger2003:nature426:154,willingale2004:mnras349:31,frail2005:apj619:994}.
In the {\it Swift} era such studies are generally limited to part of the parameter set, since they often use only one waveband, 
and therefore lack the possibility of finding the characteristic break frequencies in the broadband spectrum. 

Here, we fit the broadband spectral energy distributions (SEDs, from nIR through X-ray) of a subsample 
of the {\it BeppoSAX} sample of GRB afterglows. We constrain a subset of the blastwave parameters, 
namely the index of the power law energy distribution of electrons, $p$, the density profile parameter 
of the circumburst medium, $k$, and for some bursts the position of the cooling break, \nuc. 
Because of the paucity of radio data for most bursts in our sample, we have not included these, 
ensuring a more homogeneous approach. Radio data are not needed to determine
$p$ and $k$, since this is possible from only the nIR through X-ray SEDs.

We make use of simultaneous fits in count space to obtain the most accurate possible measurements. 
In Paper I we provide details of the observations, data reduction and fitting method, 
summarized here in Sections 2 and 3 together with the description of the model used. 
In Section 4 we present the results of our $p$- and $k$-value and cooling break analysis, 
both for the sample and for individual sources. 
We compare these results to those of previous studies of this kind in Section 5, 
and perform statistical modeling of the $p$-value dataset. 
Here we discuss our findings in the context of the fireball model and long GRB progenitors. 
We conclude by summarizing our results in Section 6.


\section{Observations and Spectral Fitting Method}

X-ray observations were made with the narrow field instruments on-board {\it BeppoSAX} (Paper I,  Table 2), 
and here we have combined data from the MECS2 and MECS3 instruments (except in the case of GRB\,970228, 
where we use the MECS3 instrument only).

Optical and nIR photometric points were taken from the literature (Paper I, Table 3) 
and from our own observations of GRB\,990510 (Curran et al. in preparation). 
Temporal decay slopes were again taken from the literature (see Paper I, Table 1): 
the optical temporal slopes from \citet{zeh2006ApJ:637:889} and the X-ray temporal slopes 
from \citet{gendre2005:aa430:465,intzand1998:apj505:L119,nicastro1999:a&as:138:437}. 
We have transformed the time of the optical and nIR photometry to the log of the midpoint of the combined X-ray observations. 
We avoid using data taken before 0.1 days after the GRB, hence we assume no complex and flaring behavior occurs 
in the lightcurves and we restrict ourselves to the slow cooling regime. 
All data are transformed to count space for fitting purposes, in order that no model need be assumed {\it a priori} 
for the X-ray spectrum, and fitted within the {\small ISIS} spectral fitting package \citep{houck2000:asp216:591}. 
Models consist of either a single or a broken power law, to allow for a possible cooling break 
in between the optical and X-rays. In the broken power law model we fix the difference in spectral slope 
to $0.5$, as expected in the case of such a cooling break. 
Both Galactic and intrinsic absorption are components in the models, allowing for either Milky Way, 
Large Magellanic Cloud (LMC) or Small Magellanic Cloud (SMC) extinction laws for the GRB host galaxy, 
at either Solar (Z$_\odot$), LMC ($Z$=1/3 Z$_\odot$) or SMC ($Z$=1/8 Z$_\odot$) metallicity. 
All details of the observations, reduction and analysis are given in Paper I, together with the results 
of the power law plus extinction fits.


\section{Theoretical Modeling}
We assume that the ambient medium density as a function of radius can be described as a power law with index $k$, 
i.e. $n(r) \propto r^{-k}$, so that a homogeneous medium is given by $k=0$ and a stellar wind environment by $k=2$ 
--- the two most likely scenarios. 
For the purpose of looking at optical and X-ray data at $\sim$ 1 day, we need to derive the time dependency 
of the peak flux $F_{\nu,max}$ and the cooling frequency $\nu_c$ as a function of $k$, 
and the peak frequency $\nu_m$ (which has no dependence on $k$), assuming the afterglows are in the slow cooling regime. 
After the jet break all of these parameters have no dependence on $k$; 
in this region we know we are dealing with a jet geometry hence we label this
case `Jet' or $_j$ (Table \ref{table:peakcooling}). 
\clearpage
\begin{table}
\begin{center}
\renewcommand{\arraystretch}{2.0}
\begin{tabular}{|l|ccc|c|} \hline
$ $ & $k$ & $k=0$ & $k=2$ & Jet \\ \hline \hline
$F_{\nu,max}$ & $-\frac{k}{2(4-k)}$ & $0$ & $-\frac{1}{2}$ & $-1$ \\ \hline
$\nu_c$ & $-\frac{4-3k}{2(4-k)}$ & $-\frac{1}{2}$ & $\frac{1}{2}$ & $0$ \\ \hline
$\nu_m$ & $-\frac{3}{2}$ & $-\frac{3}{2}$ & $-\frac{3}{2}$ & $-2$ \\ \hline
\end{tabular}
\caption{The temporal power law indices of the peak flux $F_{\nu,max}$, the cooling frequency $\nu_c$ 
and the peak frequency $\nu_m$, as a function of the circumburst density profile index $k$ for pre- (columns 2--4) and post- jet-break (column 5).}
\label{table:peakcooling}
\end{center}
\end{table}
\clearpage
If one assumes that the flux is a power law in frequency and time with $\beta$ (or $\Gamma$) the spectral slope 
and $\alpha$ the temporal slope, using the conventions 
$F_{\nu} \propto \nu^{\;-\beta} t^{\;-\alpha} \propto \nu^{\;-(\Gamma-1)} t^{\;-\alpha}$, 
with power law photon index $\Gamma = \beta + 1$, one can derive these slopes as a function of $k$ 
and the power-law index $p$ of the electron energy distribution. 
These values for $\alpha$, $\beta$ and $\Gamma$ are given in Table \ref{table:alphabeta} for two different situations: 
the observing frequency in between $\nu_m$ and $\nu_c$, and the observing frequency above both frequencies. 
Also shown in this table are the closure relations between $\alpha$ and $\beta$. 

One can invert all these relations to obtain $p$ from $\alpha$ and $\beta$, and even determine $k$ from these observables: 
\begin{equation}k\,=\,\frac{4(3\beta-2\alpha)}{3\beta-2\alpha-1}\,=\,\frac{4[3(\Gamma-1)-2\alpha]}{3\Gamma-2\alpha-4}\label{eq:kalphabeta}.\end{equation}
\clearpage
\begin{table*}
\begin{center}
\renewcommand{\arraystretch}{2.0}
\begin{tabular}{|c|cccc|cc|} \hline
 &  \multicolumn{4}{c|}{$\nu_m<\nu<\nu_c$} & \multicolumn{2}{c|}{$\nu_m<\nu_c<\nu$} \\
$$ & $k$ & $k=0$ & $k=2$ & Jet & $k$ & Jet\\ \hline \hline
$\beta$($p$) & $\frac{p-1}{2}$ & $\frac{p-1}{2}$ & $\frac{p-1}{2}$ & $\frac{p-1}{2}$ & $\frac{p}{2}$ & $\frac{p}{2}$ \\ \hline
$\Gamma$($p$) & $\frac{p+1}{2}$ & $\frac{p+1}{2}$ & $\frac{p+1}{2}$ & $\frac{p+1}{2}$ & $\frac{p+2}{2}$ & $\frac{p+2}{2}$ \\ \hline
$\alpha$($p$,$k$) & $\frac{12(p-1)-k(3p-5)}{4(4-k)}$ & $\frac{3(p-1)}{4}$ & $\frac{3p-1}{4}$ & $p$ & $\frac{3p-2}{4}$ & $p$ \\ \hline
$\alpha$($\beta$,$k$) & $\frac{6\beta(4-k)+2k}{4(4-k)}$ & $\frac{3\beta}{2}$ & $\frac{3\beta+1}{2}$ & $2\beta+1$ & $\frac{3\beta-1}{2}$ & $2\beta$ \\ \hline
$\alpha$($\Gamma$,$k$) & $\frac{6\Gamma(4-k)-8(3-k)}{4(4-k)}$ & $\frac{3(\Gamma-1)}{2}$ & $\frac{3\Gamma-2}{2}$ & $2\Gamma-1$ & $\frac{3\Gamma-4}{2}$ & $2(\Gamma-1)$ \\ \hline \hline
$p$($\beta$) & $2\beta+1$ & $2\beta+1$ & $2\beta+1$ & $2\beta+1$ & $2\beta$ & $2\beta$ \\ \hline
$p$($\Gamma$) & $2\Gamma-1$ & $2\Gamma-1$ & $2\Gamma-1$ & $2\Gamma-1$ & $2(\Gamma-1)$ & $2(\Gamma-1)$ \\ \hline
$p$($\alpha$,$k$) & $\frac{4\alpha(4-k)+12-5k}{3(4-k)}$ & $\frac{4\alpha+3}{3}$ & $\frac{4\alpha+1}{3}$ & $\alpha$ & $\frac{2(2\alpha+1)}{3}$ & $\alpha$ \\ \hline
\end{tabular}
\caption{The temporal and spectral slopes of the flux, $\alpha$ and $\beta$ (or $\Gamma$, where $\Gamma=\beta+1$) 
respectively, the closure relations between $\alpha$ and $\beta$ (or $\Gamma$), and $p$ as a function of $\alpha$, 
$\beta$ and $\Gamma$.}
\label{table:alphabeta}
\end{center}
\end{table*}
\clearpage
From Table \ref{table:alphabeta} it is clear that when the observing frequency is higher than $\nu_m$ and $\nu_c$, 
both $\alpha$ and $\beta$ only depend on $p$, and do not depend on $k$. 
In the situation where the observing frequency is situated in between $\nu_m$ and $\nu_c$, 
the spectral slope only depends on $p$, but the temporal slope depends on both $p$ and $k$. 
So the structure of the ambient medium can only be determined in the latter case ($\nu_m<\nu<\nu_c$), 
although having more accurate information on $p$ from the situation with $\nu_m<\nu_c<\nu$ is useful 
to get a better handle on $k$.


\section{Results} 

The results of fits to the SEDs for all GRBs in the sample are given in both
Table 4 and Fig. 2 of Paper I. 
For derivation of the blastwave parameter $p$ we adopt the best-fitting models as listed in Paper I 
and in particular cases additional models were included. 
SMC-like absorption was the preferred extinction model in all cases except for GRB\,000926 
where LMC-like absorption is preferred. 
We calculate the values for $p$ and $k$ for two cases: the cooling frequency in between the optical and X-rays, 
and the cooling frequency above both. We have checked whether the cooling frequency could lie below the optical band 
using the relations of the fireball model, but this turns out not to be the case for these GRBs. 
The cooling frequency, $\nu_c$, is obtained directly from the SED fits for 3 bursts: GRBs 990123, 990510 and 010222, 
with $\nu_{\mathrm{c}}$ = ($1.3^{+4.5}_{-0.9}$)$\times10^{17}$,
($4.3\pm0.5$)$\times10^{15}$, and ($4.1^{+15.2}_{-4.1}$)$\times 10^{15}$ Hz 
at 1.245, 1.067 and 1.511 days since burst respectively. Applying the fireball model we find that another two sources, 
GRBs 980329 and 980703, may require a break within their SEDs, at $\sim
2.6\times10^{17}$ and $\sim 8\times10^{17}$ Hz at 1.148 and 1.333 days since burst
(the logarithmic midpoint of the X-ray spectrum) respectively.

The resulting values for $p$ can be found in Table \ref{table:pvalues} and for $k$ in Table \ref{table:kvalues}. 
All errors are given at the 90$\%$ confidence level for one interesting parameter, unless otherwise stated. 
$\alpha_1$ and $\alpha_2$ refer to the pre- and post-break optical lightcurve slopes given in Paper I; 
we allowed for the possibility that these breaks are not jet breaks by considering that $\alpha_2$ 
is both pre- (columns 6 and 7) and post- (column 8) jet break. This has also been done for the X-ray temporal slopes.

\subsection{Results: Individual Sources}

\subsubsection{GRB\,970228}

We find $p\,=\,2.44_{-0.06}^{+0.18}$ and $k\,=\,1.73_{-1.69}^{+0.56}$, with $\nu_{\mathrm{X}}<\nu_{\mathrm{c}}$ at $0.52$ days. 
A cooling break between optical and X-ray bands is not required at the time of
our SED ($0.52$ days) according to the F-test: the F-test probability, the probability that 
the result is obtained by chance, is $2.1\times 10^{-2}$, which is quite high; so adding one
extra free parameter is not a significant improvement. 
Using the best fitting model of a single power law plus SMC-like extinction,
in the regime $\nu_{\mathrm{X}}<\nu_{\mathrm{c}}$, we find that the data can
be fit by both a homogeneous and a wind-like circumburst medium; 
the value of $k$ is best constrained by the optical temporal slope. 

\subsubsection{GRB\,970508}

We find $p\,=\,2.56_{-0.46}^{+0.10}$ and $k\,=\,0.49_{-0.67}^{+1.36}$, with $\nu_{\mathrm{X}}<\nu_{\mathrm{c}}$ at $1.68$ days. 
Using the single power law with LMC extinction (rather than SMC to obtain the
more conservative errors on the spectral slope) and optical to X-ray offset free, in the regime
$\nu_{\mathrm{X}}<\nu_{\mathrm{c}}$, we find the data are best fit by a homogeneous medium; 
the value $k$ is best constrained by the optical temporal slope. 
Broadband modeling by Van der Horst \& Wijers (in preparation) with $k$ as a free parameter gives very tight constraints on $k$: 
a value of $0.02\pm 0.03$ is derived, i.e. a homogeneous medium.

Previous works put the cooling break at optical frequencies, $\nu_c\,=\,1.6\times 10^{14}$ Hz, at $12.1$ days since burst, 
between the $B$ and $V$ bands \citep[e.g.][]{wijers1999:apj523:177}. 
We, however, find that the cooling break is likely to lie above the X-rays at 1.68 days. 
We note in this context the uncertain extrapolation of the optical data used in the
SED, owing to an irregular shaped lightcurve at early times, which we have
attempted to account for in allowing the optical to X-ray offset to go
free. This is a particularly difficult case given that the X-ray data cover
the time period immediately following an optical flare when the optical
lightcurve appears to have flattened before breaking to a typical and
well defined power law.

\subsubsection{GRB\,971214}

We find $p\,=\,2.20\pm0.06$ and $k\,=\,2.17_{-0.35}^{+0.25}$, with $\nu_{\mathrm{X}}<\nu_{\mathrm{c}}$ at $1.36$ days. 
Using the best-fitting model of a power law plus SMC extinction, in the
regime $\nu_{\mathrm{X}}<\nu_{\mathrm{c}}$, the data are best fit by a wind
medium: from the optical temporal slope we find $k\,=\,2.17_{-0.35}^{+0.25}$, whilst from the X-ray temporal slope
$k\,=\,2.33_{-0.34}^{+0.24}$. A broken power law does not provide a significant improvement in the fit compared to a single power law, 
i.e. the F-test probability is high ($4.1\times 10^{-2}$). 
A spectral break is claimed for this burst in the IR ($\sim$1 micron) at 0.58 days \citep[$\nu_c \sim
3 \times 10^{14}$ Hz, ][]{ramaprakash1998:Natur393:43}. This is not the cooling break, but the peak of the SED 
moving to lower frequencies, so there is no conflict with our results for $\nu_{\mathrm{c}}$.

\subsubsection{GRB\,980329}

We find $p\,=\,2.50_{-0.62}^{+0.20}$ and $k\,=\,-4.89_{-1.40}^{+25.98}$, with $\nu_{\mathrm{c}}<\nu_{\mathrm{X}}$ at $1.15$ days. 
The spectral fit obtained with a single power law plus SMC extinction is inconsistent with the optical temporal slopes. 
A spectral break in the power law does not provide a significant improvement of the fit according to the F-test (probability of $7.2\times 10^{-2}$), 
but this spectral break model provides agreement between the spectral slopes and the optical and
X-ray temporal slopes. In this regime, $\nu_{\mathrm{c}}<\nu_{\mathrm{X}}$, $k$ cannot be well constrained though
the centroid of the fit to the optical data is that of a homogeneous medium. 
We note that when omitting the $I$ band point from the SED, which may be
overestimated \citep[see][ and Paper I]{yost2002:ApJ577:155}, our results do not change.

\subsubsection{GRB\,980519}

We find $p\,=\,2.96_{-0.08}^{+0.06}$ and $k\,=\,0.23_{-3.05}^{+1.22}$, with $\nu_{\mathrm{X}}<\nu_{\mathrm{c}}$ at $0.93$ days. 
Using the power law plus SMC extinction model for the SED, in the regime
$\nu_{\mathrm{X}}<\nu_{\mathrm{c}}$, the optical data are best fit by
a homogeneous medium, and the X-ray temporal slopes can be fitted by both a
homogeneous and a wind medium; $k$ is therefore best constrained by the optical temporal slope. 
In contrast, \citet{chevalier1999:apj520:29} found that the radio emission of the afterglow of GRB 980519 measured between 7.2 hours and 63 days
since the burst is consistent with an external wind instead of a homogeneous medium. \citet{frail2000:apj534:559} note, however, that the 
interstellar scintillation present in the radio data does not allow one to draw firm conclusions on this. 

The optical temporal break at $0.48\pm 0.03$ days \citep{zeh2006ApJ:637:889} cannot be
explained by passage through the optical bands of $\nu_{\mathrm{c}}$, since the derived value for $p$ 
from the temporal slope is too high in that case ($p\,=\,3.69\pm 0.06$) compared to the $p$-value from the joint spectral slope. 
It also cannot be explained by a jet break, since $p$ is too low in that case ($p\,=\,2.27\pm 0.05$).
It appears that the fireball model is a good explanation for the first
temporal slope and the spectrum used here, but the post-break optical slope has
either been incorrectly measured or we do not yet have the correct model for
this afterglow. 
We note that this afterglow showed a very steep temporal decay compared to other GRBs
\citep{halpern1999:ApJL517:105}. This is somewhat reminiscent of the very
early-time decays of many {\it Swift} bursts which occur at $\le$500 s after
trigger \citep{nousek2006:apj642:389} and likely have a significant prompt emission component.

\subsubsection{GRB\,980703}

We find $p\,=\,2.74_{-0.48}^{+0.10}$ and $k\,=\,1.63_{-56.46}^{+1.34}$, with $\nu_{\mathrm{c}}<\nu_{\mathrm{X}}$ at $1.33$ days. 
The spectral fit obtained with a single power law plus SMC extinction is inconsistent with the X-ray temporal slopes. 
A spectral break in the power law does not provide a significant improvement of the fit according to the F-test (probability of $7.2\times 10^{-2}$), 
but this spectral break model provides agreement between the spectral slopes and the optical and X-ray temporal slopes. 
Hence we use the broken power law plus SMC extinction model in the regime $\nu_{\mathrm{c}}<\nu_{\mathrm{X}}$. 
The nature of the optical temporal break at $1.35\pm 0.94$ days cannot be
determined because of large uncertainties in the optical temporal slopes,
which are also the reason why $k$ cannot be constrained.

Two publications have postulated a position for the cooling break in past
studies. \citet{vreeswijk1999:ApJ523:171} propose $\nu_{\mathrm{o}} < \nu_{\mathrm{X}} < \nu_{\mathrm{c}}$ at $1.2$ days after the burst, 
\citet{bloom1998:apj508} propose $\nu_{\mathrm{o}} < \nu_{\mathrm{c}} < \nu_{\mathrm{X}}$ at $5.3$ days,  
and our SED study at $1.33$ days, when compared with the optical temporal slope,
is inconclusive since both $\nu_{\mathrm{X}} < \nu_{\mathrm{c}}$ and $\nu_{\mathrm{X}} > \nu_{\mathrm{c}}$ can be
accommodated. It may be that the cooling break has moved into our observed
bands during accumulation of the X-ray spectrum (possibly indicated by the
inability of the fireball model to fit the data when a single power law is
assumed for the spectrum). If we require consistency with these previous results, the
cooling break must be moving to lower energies and lies approximately at X-ray
frequencies in our data. This would mean that the circumburst medium is homogeneous, since
$\nu_{\mathrm{c}}$ is expected to move as $t^{-1/2}$ in this case, 
while $\nu_{\mathrm{c}}$ will increase in time as $t^{1/2}$ in the wind case.

The host galaxy of GRB\,980703 appears to have a high and possibly variable
optical extinction along the line of sight to the GRB (see Paper
I). The different (and formally inconsistent) values of $A_{V}$ may be due to
different methods for measuring the extinction, probing of different regions of
the host galaxy, or may indicate that the environment in which the burst occurred
is changing with time. We have used the \citet{vreeswijk1999:ApJ523:171} optical data
and scaled it from 1.2 days to 1.33 days after trigger. Combining the optical
and X-ray data when fitting provides us with a different estimate for the
extinction than was obtained by \citet{vreeswijk1999:ApJ523:171} for the optical alone. Any
change in optical extinction will have an effect upon the measured spectral
slope and hence the derived value of $p$.

\subsubsection{GRB\,990123}

We find $p\,=\,1.99_{-0.07}^{+0.00}$ and $k\,=\,2.00_{-0.21}^{+0.26}$, with $\nu_{\mathrm{c}}<\nu_{\mathrm{X}}$ at $1.25$ days. 
With this best fitting SED model of a broken power law plus SMC extinction,
in the regime $\nu_{\mathrm{c}}<\nu_{\mathrm{X}}$, the optical temporal decay is best fit by a wind medium. 
The optical temporal break at $2.06\pm 0.83$ days is marginally consistent
with a jet-break: the $p$-value derived from the post-break temporal slope is consistent with the one derived from the pre-break optical temporal slope, 
but inconsistent with the spectral slope. The uncertainties in the X-ray temporal slope are too large to determine the phase of blastwave evolution
(i.e. before or after jet-break) from the X-ray data alone.

For GRB\,990123 the temporal slope difference between optical and X-ray of 0.25
also agrees with the spectral analysis where a broken power
law model is the best fit, indicating a cooling break between the optical and
X-ray bands at 1.25 days post-burst. The value we derive for $p$ is consistent with that derived from the X-ray
spectrum alone of $p\,=\,2.0\pm0.1$ \citep{stratta2004:apj608:846} and lower than a
previous estimate via broadband modeling of $p\,=\,2.28\pm0.05$
\citep{panaitescu2002:apj571} (we note that the latter uncertainty is $1\,\sigma$ and not the 90\%
error used in the rest of our paper).

\subsubsection{GRB\,990510}

We find $p\,=\,2.06_{-0.02}^{+0.14}$ and $k\,=\,0.80_{-0.90}^{+0.22}$, with $\nu_{\mathrm{c}}<\nu_{\mathrm{X}}$ at $1.07$ days. 
The best fitting model to the SED is clearly a broken power law with negligible extinction,
in the regime $\nu_{\mathrm{c}}<\nu_{\mathrm{X}}$. There is considerable improvement in the $\chi^2$ when allowing for this break in
the power law, noted by previous authors, which we find is located at 0.016 -
0.020 keV at $\sim$1.07 days since burst \citep[consistent with the value 
given by][]{pian2001:AandA372:456}. The slope change is as expected for a cooling break in the slow
cooling regime when leaving both power law slopes free. 

The optical data are best fit by the values for $p$ and $k$ mentioned above. 
In this case, however, the X-ray temporal slope
is not consistent with the spectral slope nor the optical temporal slope at
the 90\% ($\sim 1.6\sigma$) level, but does agree within $3\,\sigma$. The 
optical temporal break at $1.31\pm 0.07$ days is marginally consistent with a
jet-break. The derived value of \p\ is consistent with the value derived from the {\it BeppoSAX} X-ray spectrum
alone of $p\sim2.1$, by \citet{kuulkers2000:ApJ:538:638}, and rules out the value of
$p=2.6\pm0.2$ also derived from the {\it BeppoSAX} X-ray spectrum by \citet{stratta2004:apj608:846}; 
that same X-ray data is used here, but is combined with nIR and optical data to obtain our limits on \p.

\subsubsection{GRB\,000926}

We find $p\,=\,2.54_{-0.08}^{+0.14}$ and $k\,=\,2.16_{-0.30}^{+0.17}$, with $\nu_{\mathrm{X}}<\nu_{\mathrm{c}}$ at $2.23$ days. 
With the power law plus LMC extinction model, in the regime $\nu_{\mathrm{X}}<\nu_{\mathrm{c}}$, the optical temporal decay is best fit by a wind-like medium. 
Large uncertainties in the X-ray temporal slope prevent determination of the
circumburst medium structure and blastwave evolution phase from X-ray data alone; the 
optical temporal break at $2.10\pm 0.15$ days is consistent with a jet-break.

\subsubsection{GRB\,010222}

We do not find a consistent solution for this afterglow taking $90\%$ uncertainties, 
but we do find one taking $3\sigma$ uncertainties: $p\,=\,2.04_{-0.10}^{+0.18}$ and $k\,=\,2.28_{-0.29}^{+0.15}$, 
with $\nu_{\mathrm{c}}<\nu_{\mathrm{X}}$ at $1.51$ days. 
Adopting the single power law model with LMC extinction, in the regime $\nu_{\mathrm{X}}<\nu_{\mathrm{c}}$, 
the optical slopes are not consistent with the spectral slope nor the X-ray
temporal slopes, both at the $90\%$ and $3\sigma$ levels. 
Using a broken power law plus SMC extinction, in the regime $\nu_{\mathrm{c}}<\nu_{\mathrm{X}}$, 
for which the F-test indicates a marginal improvement (probability of $1\times 10^{-4}$), 
the X-ray temporal slope is not consistent with the spectral slope nor the
optical temporal slope at the $90\%$ level, 
but they are consistent at $3\sigma$. 
In the first case we obtain $p\,=\,2.72_{-0.05}^{+0.05}$ and $k\,=\,0.30_{-0.84}^{+0.59}$; 
in the latter case $p\,=\,2.04_{-0.10}^{+0.18}$ and $k\,=\,-2.25_{-43}^{+2.54}$, which is derived from the pre-break optical temporal slope, 
or $k\,=\,2.28_{-0.29}^{+0.15}$, derived from the post-break optical temporal slope. 
Since the temporal break happens quite early, $0.64\pm0.09$ days after the burst \citep{zeh2006ApJ:637:889}, 
and the post-break optical slopes are inconsistent with a jet-interpretation, 
the early temporal slope is probably influenced by late-time energy injection 
and a medium with $k\,=\,2.28_{-0.29}^{+0.15}$ the correct interpretation. 

\citet{panaitescu2002:apj571} find the cooling
break to lie at optical wavelengths or longer, in agreement with our
results. However, they derive a low value for $p$ of $1.35$, and find
significant reddening of the optical spectrum which they say explains the
second steepening observed in the optical after 6 days by
\citet{fruchter2001GCN1087:1}. 
\citet{bhattacharya2004:romeconf:411} obtain good fits by adopting a two-slope electron energy distribution with $p$-values of $1.3$ and $2.1$, 
below and above a so-called injection break, respectively. This injection break is located in the X-ray regime at $\sim 1$ days after the burst. 
In their model the optical temporal break is a jet-break and the circumburst medium is homogeneous. 
\citet{bjornsson2002:apj579} argue that the unusually slow decay of this afterglow
and positive detection of linear polarization can be explained by a jet model
with continuous energy injection. Such slow decays are seen in the `canonical'
{\it Swift} X-ray lightcurves \citep[e.g.][]{nousek2006:apj642:389}, termed the plateau phase, and typically
begin a few hundred seconds after the GRB trigger. The electron energy distribution then has $p\,=\,2.49\pm0.05$, 
which is inconsistent with our result for $p$ from the spectral fits.


\section{Discussion}

\begin{figure}
\begin{center}
\includegraphics[width=10cm, angle=0]{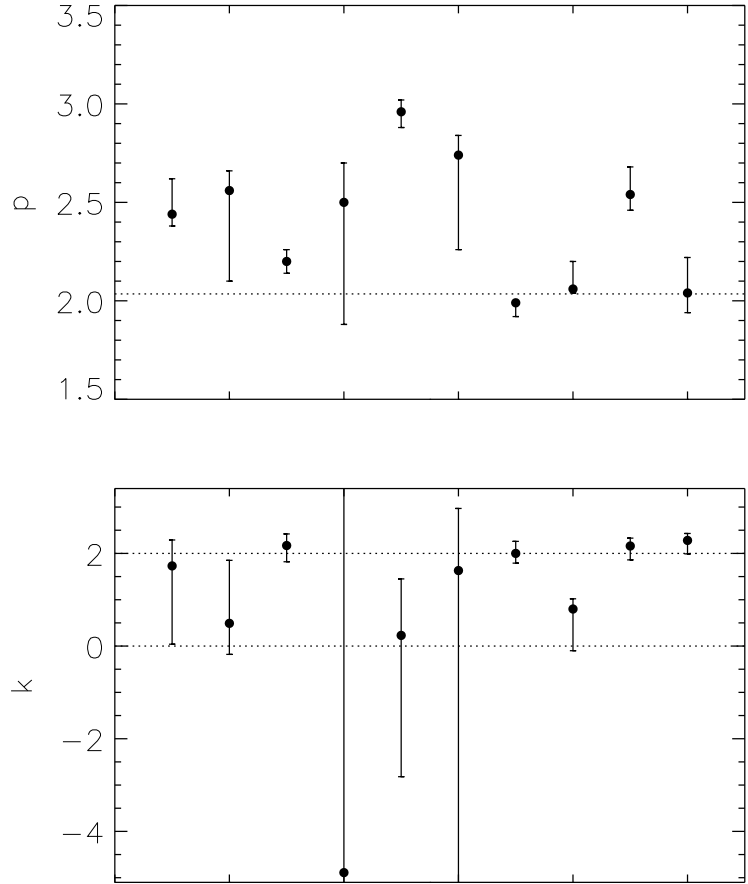}
\caption{Derived values of $p$ (top panel) and $k$ (lower panel) for each individual afterglow 
(see Table \ref{table:finalresults}): the horizontal axes represent the GRBs in date order left to right 
and errors are $90\,\%$ confidence. In the top panel the dotted line indicates the most likely value of $p=2.04$; 
in the lower panel the dotted lines indicate $k=0$ (homogeneous medium) and $k=2$ (stellar wind).} 
\label{pvalues}
\end{center}
\end{figure}

The parameters that can be derived from broadband modeling of afterglow lightcurves describe the micro- 
and macrophysics of the relativistic jet and its surrounding medium. To obtain the full set of parameters 
the spectral energy distribution has to be covered from X-ray to radio wavelengths. Two of these 
parameters can be deduced from just the spectral and temporal slopes in the optical and X-rays, i.e. the 
electron energy distribution index $p$ and the circumburst medium profile parameter $k$. These two parameters 
have been determined in this paper for a selection of 10 GRBs, for which the final results are shown in Table 
\ref{table:finalresults} and Fig. \ref{pvalues}. 
For completeness the values for $\nu_{\mathrm{c}}$ are also listed in Table \ref{table:finalresults}; 
half of the GRBs in this sample have a value $\nu_{\mathrm{c}}\le \nu_{\mathrm{X}}\sim 8\times10^{17}$ Hz. 
\clearpage
\thispagestyle{empty}
\setlength{\voffset}{20mm}
\begin{deluxetable}{c |c c| c c c c c| c c c c c|}
\tabletypesize{\scriptsize}
\renewcommand{\arraystretch}{1.5}
\rotate
\tablecaption{Values for $p$. We calculate the results for the cases $k=0$ and $k=2$. 
Bold type highlights consistent results (at the $90\%$ level). In cases where the best-fitting spectral model to the SED (from Paper I) 
is inconsistent with the model fits, we show the results for this original best-fitting model in italics.
\label{table:pvalues}}
\tablewidth{0pt}
\tablehead{ & \multicolumn{2}{c|}{Spectral} & \multicolumn{5}{c|}{Optical Temporal}&\multicolumn{5}{c|}{X-ray Temporal} \\
\colhead{GRB} & \colhead{$p\;$($\Gamma$)} & \colhead{$p\;$($\Gamma$)} & \colhead{$p\;$($\alpha_1,0$)} & \colhead{$p\;$($\alpha_1,2$)} & \colhead{$p\;$($\alpha_2,0$)} & \colhead{$p\;$($\alpha_2,2$)} & \colhead{$p\;$($\alpha_2,j$)} & \colhead{$p\;$($\alpha_x,0$)} & \colhead{$p\;$($\alpha_x,2$)} & \colhead{$p\;$($\alpha_x,j$)} & \colhead{$p\;$($\alpha_x$)} & \colhead{$p\;$($\alpha_x,j$)} \\ 
$$ & $\nu_{\mathrm{X}}<\nu_{\mathrm{c}}$ & $\nu_{\mathrm{c}}<\nu_{\mathrm{X}}$
&  &  &  &  &  & $\nu_{\mathrm{X}}<\nu_{\mathrm{c}}$ &
$\nu_{\mathrm{X}}<\nu_{\mathrm{c}}$ & $\nu_{\mathrm{X}}<\nu_{\mathrm{c}}$ &
$\nu_{\mathrm{c}}<\nu_{\mathrm{X}}$ & $\nu_{\mathrm{c}}<\nu_{\mathrm{X}}$ 
}
\startdata
970228 & {\bf 2.44}$_{-0.06}^{+0.18}$ & $2.12_{-0.06}^{+0.46}$ & ${\bf 2.95\pm 0.32}$ & ${\bf 2.28\pm 0.32}$ & ... & ... & ... & ${\bf 2.73\pm 0.43}$ & ${\bf 2.07\pm 0.43}$ & $1.30\pm 0.32$ & $2.40\pm 0.43$ & $1.30\pm 0.32$ \\ 
970508 & {\bf 2.56}$_{-0.46}^{+0.10}$ & $2.28_{-0.58}^{+0.28}$ & ${\bf 2.65\pm 0.02}$ & $1.99\pm 0.02$ & ... & ... & ... & ${\bf 2.47\pm 0.21}$ & $1.80\pm 0.21$ & $1.10\pm 0.16$ & $2.13\pm 0.21$ & $1.10\pm 0.16$ \\ 
971214 & {\bf 2.20}$_{-0.06}^{+0.06}$ & $2.08_{-0.28}^{+0.18}$ & $2.99\pm 0.17$ & ${\bf 2.32\pm 0.17}$ & ... & ... & ... & $3.13\pm 0.21$ & ${\bf 2.47\pm 0.21}$ & $1.60\pm 0.16$ & $2.80\pm 0.21$ & $1.60\pm 0.16$ \\ 
980329 & {\it 2.64}$_{-0.14}^{+0.08}$ & {\bf 2.50}$_{-0.62}^{+0.20}$ & ${\bf 2.13\pm 0.26}$ & $1.47\pm 0.26$ & ... & ... & ... & ${\it 3.00\pm 0.43}$ & ${\it 2.33\pm 0.43}$ & $1.50\pm 0.32$ & ${\bf 2.67\pm 0.43}$ & $1.50\pm 0.32$ \\ 
980519 & {\bf 2.96}$_{-0.08}^{+0.06}$ & $2.86_{-0.20}^{+0.14}$ & ${\bf 3.00\pm 0.26}$ & $2.33\pm 0.26$ & $4.03\pm 0.06$ & $3.36\pm 0.06$ & $2.27\pm 0.05$ & ${\bf 3.44\pm 0.64}$ & ${\bf 2.77\pm 0.64}$ & $1.83\pm 0.48$ & $3.11\pm 0.64$ & $1.83\pm 0.48$ \\ 
980703 & {\it 2.84}$_{-0.06}^{+0.06}$ & {\bf 2.74}$_{-0.48}^{+0.10}$ & ${\bf 2.13\pm 1.79}$ & ${\bf 1.47\pm 1.79}$ & ${\bf 3.20\pm 0.98}$ & ${\bf 2.53\pm 0.98}$ & ${\bf 1.65\pm 0.74}$ & $2.20\pm 0.43$ & $1.53\pm 0.43$ & $0.90\pm 0.32$ & ${\bf 1.87\pm 0.43}$ & $0.90\pm 0.32$ \\ 
990123 & $2.22_{-0.02}^{+0.02}$ & {\bf 1.99}$_{-0.07}^{+0.00}$ & $2.65\pm 0.13$ & ${\bf 1.99\pm 0.13}$ & $3.16\pm 0.32$ & $2.49\pm 0.32$ & ${\bf 1.62\pm 0.24}$ & $3.75\pm 1.77$ & $3.08\pm 1.77$ & $2.06\pm 1.33$ & ${\bf 3.41\pm 1.77}$ & ${\bf 2.06\pm 1.33}$ \\ 
990510 & $2.71_{-0.02}^{+0.01}$ & {\bf 2.06}$_{-0.02}^{+0.14}$ & ${\bf 2.23\pm 0.04}$ & $1.56\pm 0.04$ & $3.80\pm 0.13$ & $3.13\pm 0.13$ & ${\bf 2.10\pm 0.10}$ & $2.87\pm 0.21$ & $2.20\pm 0.21$ & $1.40\pm 0.16$ & $2.53\pm 0.21$ & $1.40\pm 0.16$ \\ 
000926 & {\bf 2.54}$_{-0.08}^{+0.14}$ & $2.50_{-0.56}^{+0.16}$ & $3.32\pm 0.06$ & ${\bf 2.65\pm 0.06}$ & $4.27\pm 0.11$ & $3.60\pm 0.11$ & ${\bf 2.45\pm 0.08}$ & ${\bf 3.27\pm 1.07}$ & ${\bf 2.60\pm 1.07}$ & ${\bf 1.70\pm 0.80}$ & $2.93\pm 1.07$ & $1.70\pm 0.80$ \\ 
010222 & {\it 2.72}$_{-0.05}^{+0.05}$ & {\bf 2.04}$_{-0.10}^{+0.18}$ & ${\bf 1.80\pm 0.19}$ & $1.13\pm 0.19$ & $2.92\pm 0.04$ & ${\bf 2.25\pm 0.04}$ & $1.44\pm 0.03$ & ${\it 2.77\pm 0.09}$ & $2.11\pm 0.09$ & $1.33\pm 0.06$ & $2.44\pm 0.09$ & $1.33\pm 0.06$ \\ 
\enddata
\end{deluxetable}
\clearpage
\setlength{\voffset}{0mm}
\begin{table*}
\begin{center}
\renewcommand{\arraystretch}{2.0}
\begin{tabular}{|c||cc|c||c|c|} \hline
GRB & $k\;$($\alpha_1,\Gamma$) & $k\;$($\alpha_2,\Gamma$) & $k\;$($\alpha_x,\Gamma$) & $k\;$($\alpha_1,\Gamma$) & $k\;$($\alpha_2,\Gamma$) \\ 
$$ & $\nu_{\mathrm{X}}<\nu_{\mathrm{c}}$ & $\nu_{\mathrm{X}}<\nu_{\mathrm{c}}$ & $\nu_{\mathrm{X}}<\nu_{\mathrm{c}}$ & $\nu_{\mathrm{c}}<\nu_{\mathrm{X}}$ & $\nu_{\mathrm{c}}<\nu_{\mathrm{X}}$ \\ \hline\hline
970228 & {\bf 1.73}$_{-1.69}^{+0.56}$ & ... & {\bf 1.22}$_{-4.77}^{+0.93}$ & $2.21_{-1.95}^{+0.36}$ & ... \\ \hline
970508 & {\bf 0.49}$_{-0.67}^{+1.36}$ & ... & {\bf -0.65}$_{-5.61}^{+2.51}$ & $1.44_{-1.05}^{+0.94}$ & ... \\ \hline
971214 & {\bf 2.17}$_{-0.35}^{+0.25}$ & ... & {\bf 2.33}$_{-0.34}^{+0.24}$ & $2.31_{-0.49}^{+0.38}$ & ... \\ \hline
980329 & {\it -12.67}$_{-43.94}^{+31.82}$ & ... & {\it 1.40}$_{-2.53}^{+0.92}$ & {\bf -4.89}$_{-1.40}^{+25.98}$ & ... \\ \hline
980519 & {\bf 0.23}$_{-3.05}^{+1.22}$ & $2.46_{-0.12}^{+0.12}$ & {\bf 1.67}$_{-3.64}^{+0.90}$ & $0.69_{-3.19}^{+1.19}$ & $2.55_{-0.18}^{+0.18}$ \\ \hline
980703 & {\it 70.67}$_{-68.24}^{+65.16}$ & {\it 1.40}$_{-22.72}^{+184.42}$ & {\it -96.00}$_{-93.22}^{+107.84}$ & {\bf -40.44}$_{-42.92}^{+46.57}$ & {\bf 1.63}$_{-56.46}^{+1.34}$ \\ \hline
990123 & $1.58_{-0.38}^{+0.29}$ & $2.34_{-0.45}^{+0.29}$ & $2.78_{-5.41}^{+0.55}$ & {\bf 2.00}$_{-0.21}^{+0.26}$ & {\bf 2.55}$_{-0.31}^{+0.25}$ \\ \hline
990510 & $-10.55_{-6.51}^{+3.71}$ & $2.48_{-0.13}^{+0.12}$ & $0.76_{-1.24}^{+0.72}$ & {\bf 0.80}$_{-0.90}^{+0.22}$ & {\bf 2.89}$_{-0.14}^{+0.06}$ \\ \hline
000926 & {\bf 2.16}$_{-0.30}^{+0.17}$ & $2.89_{-0.13}^{+0.08}$ & {\bf 2.09}$_{-12.37}^{+0.86}$ & $2.21_{-0.32}^{+0.53}$ & $2.90_{-0.13}^{+0.24}$ \\ \hline
010222 & {\it 14.53}$_{-5.12}^{+224.77}$ & {\it 0.92}$_{-0.36}^{+0.30}$ & {\it 0.30}$_{-0.84}^{+0.59}$ & {\bf -2.25}$_{-42.53}^{+2.54}$ & {\bf 2.28}$_{-0.29}^{+0.15}$ \\ \hline
\end{tabular}
\caption{Values for $k$. Bold type highlights consistent results (at the $90\%$ level). 
In cases where the best-fitting spectral model to the SED (from Paper I) is inconsistent with the model fits, 
we show the results for this original best-fitting model in italics.}
\label{table:kvalues}
\end{center}
\end{table*}

\clearpage

\clearpage
\begin{table}
\begin{center}
\renewcommand{\arraystretch}{2.0}
\begin{tabular}{|c|c|c|l|l|} \hline
GRB & $p$ & $k$ & Medium & $\nu_{\mathrm{c}}$ (Time of SED)\\ \hline
970228 & $2.44_{-0.06}^{+0.18}$ & $1.73_{-1.69}^{+0.56}$ & wind/homogeneous & $> \nu_{\mathrm{X}}\sim 8\times10^{17}$ ($0.52$ days)\\
970508\tablenotemark{a} & $2.56_{-0.46}^{+0.10}$ & $0.49_{-0.67}^{+1.36}$ & homogeneous & $> \nu_{\mathrm{X}}$ ($1.68$ days)\\
971214 & $2.20_{-0.06}^{+0.06}$ & $2.17_{-0.35}^{+0.25}$ & wind & $> \nu_{\mathrm{X}}$ ($1.36$ days)\\
980329\tablenotemark{b} & $2.50_{-0.62}^{+0.20}$ & $-4.89_{-1.40}^{+25.98}$ & wind/homogeneous &  $\sim 2.6\times 10^{17}$  ($1.15$ days)\\
980519\tablenotemark{c} & $2.96_{-0.08}^{+0.06}$ & $0.23_{-3.05}^{+1.22}$ & homogeneous & $> \nu_{\mathrm{X}}$ ($0.93$ days)\\
980703\tablenotemark{d} & $2.74_{-0.48}^{+0.10}$ & $1.63_{-56.46}^{+1.34}$ & wind/homogeneous & $\sim \nu_{\mathrm{X}}$ ($1.33$ days)\\
990123 & $1.99_{-0.07}^{+0.00}$ & $2.00_{-0.21}^{+0.26}$ & wind & $1.3^{+4.5}_{-0.9}\times 10^{17}$  ($1.25$ days)\\
990510\tablenotemark{e} & $2.06_{-0.02}^{+0.14}$ & $0.80_{-0.90}^{+0.22}$ & homogeneous & $4.3\pm0.5\times 10^{15}$  ($1.07$ days)\\
000926\tablenotemark{f} & $2.54_{-0.08}^{+0.14}$ & $2.16_{-0.30}^{+0.17}$ & wind & $> \nu_{\mathrm{X}}$ ($2.23$ days)\\
010222\tablenotemark{g} & $2.04_{-0.10}^{+0.18}$ & $2.28_{-0.29}^{+0.15}$ & wind & $4.1^{+15.2}_{-4.1}\times 10^{15}$  ($1.51$ days)\\
\hline
\end{tabular}
\caption{Final results for $p$, $k$ and $\nu_{\mathrm{c}}$ for all ten bursts in the sample.}
\label{table:finalresults}
\tablenotetext{a}{Optical data extrapolation is uncertain}
\tablenotetext{b}{Broken power law better than single power law in SED.}
\tablenotetext{c}{This solution is consistent with all measurements except the post-break optical temporal slope.}
\tablenotetext{d}{Broken power law better than single power law in SED. 
Large uncertainties in the optical temporal slopes.}
\tablenotetext{e}{X-ray temporal slope only consistent at $3\sigma$ level.}
\tablenotetext{f}{X-ray temporal slopes have large uncertainties.}
\tablenotetext{g}{Break in the optical lightcurve at 0.6 days is not a jet-break. X-ray temporal slope only consistent at $3\sigma$ level.}
\end{center}
\end{table}

\clearpage


\subsection{The Distribution of $p$}

Some theoretical studies of particle acceleration by ultra-relativistic shocks indicate 
that there is a nearly universal value of $p$ of $2.2-2.3$ \citep[e.g.][]{kirk2000:apj542:235,achterberg2001MNRAS32:393}, 
while other studies indicate that there is a large range of possible values for $p$ of $1.5-4$ 
\citep[e.g.][]{barringNucPhysB2004:136:198}. 
From the results presented in this paper and from broadband modeling of individual bursts, 
quite a large range of values for $p$ have been found, which could indicate that there is 
a large intrinsic scatter in the value of $p$. 
Here we test the null-hypothesis namely that the observed distribution of $p$ can be obtained 
from a parent distribution with a single central value of $p$, 
by performing a statistical log-likelihood analysis on the obtained values of $p$. 

We first determine the most likely value of $p$ for our sample, by minimizing the log-likelihood of our 10 measured values of $p$. 
In order to do this, we describe the measured values of $p$ and their uncertainties as probability distributions, 
in which we take asymmetric measurement uncertainties into account. 
These probability distributions are given as two halves of normal distributions with different widths that are set by the measured uncertainties. 
The two halves of the probability distributions are normalized such that they are connected as a continuous function and have a total integral equal to 1. 
This results in distribution functions which are given by
\begin{equation}P(p,p_{\mathrm{meas}},\sigma_1,\sigma_2)\,=\,\frac{\sqrt{2}}{\sqrt{\pi}(\sigma_1+\sigma_2)}\,\cdot \left\{\begin{array}{ll} e^{-(p-p_{\mathrm{meas}})^2/2\sigma_1^2} & (p<p_{\mathrm{meas}}) \\ e^{-(p-p_{\mathrm{meas}})^2/2\sigma_2^2} & (p>p_{\mathrm{meas}}) \end{array} \right. \,,\end{equation}
with $\sigma_1$ and $\sigma_2$ the lower and upper $1\,\sigma$ uncertainties in the measured value of $p$ (indicated as $p_{\mathrm{meas}}$) respectively. 
This probability function describes how likely a value of $p$ is given the measurement ($p_{\mathrm{meas}}$,$\sigma_1$,$\sigma_2$). 
To convert the $90\,\%$ confidence limits in Table \ref{table:finalresults} to $1\,\sigma$ uncertainties, 
we divided those by a factor of $1.6$. 
The log-likelihood for these probability distributions is given by
\begin{eqnarray}-2\log\left({\displaystyle\prod_{i=1}^{N}P_i}\right)&=&\displaystyle\sum_{i=1}^N\left(-2\log{P_i}\right)\nonumber \\ 
&=&N\log{(2\pi)}+\displaystyle\sum_{i=1}^N\left[\log{\left(\frac{\sigma_{1,i}+\sigma_{2,i}}{2}\right)^2+\left(\frac{p-p_{\mathrm{meas},i}}{\sigma_{1/2,i}}\right)^2}\right]\,,\end{eqnarray}
in which in the last term $\sigma_{1,i}$ is used for $p<p_{\mathrm{meas},i}$ and $\sigma_{2,i}$ for $p>p_{\mathrm{meas},i}$. 
$N$ is the number of measurements, i.e. in this case $N=10$. 

The log-likelihood is a function of $p$ and we determine the most probable $p$-value by minimizing this function for our 10 bursts. 
We find the most likely value of $p=2.04^{+0.02}_{-0.03}$, with a log-likelihood of $613.6$. 
We have derived the uncertainties in this most likely $p$-value by generating $10^5$ synthetic datasets for the 10 bursts in our sample. 
These datasets are possible measurements within the measurement uncertainties, generated by taking random numbers from the probability distributions 
that are defined by the measured $p$-values and their uncertainties. 
For each possible dataset we determine the most likely $p$-value, and from the resulting distribution of most likely $p$-values 
we obtain the $1\,\sigma$ uncertainties in the value of $p=2.04$.

\begin{figure}
\begin{center}
\includegraphics[width=10cm, angle=-90]{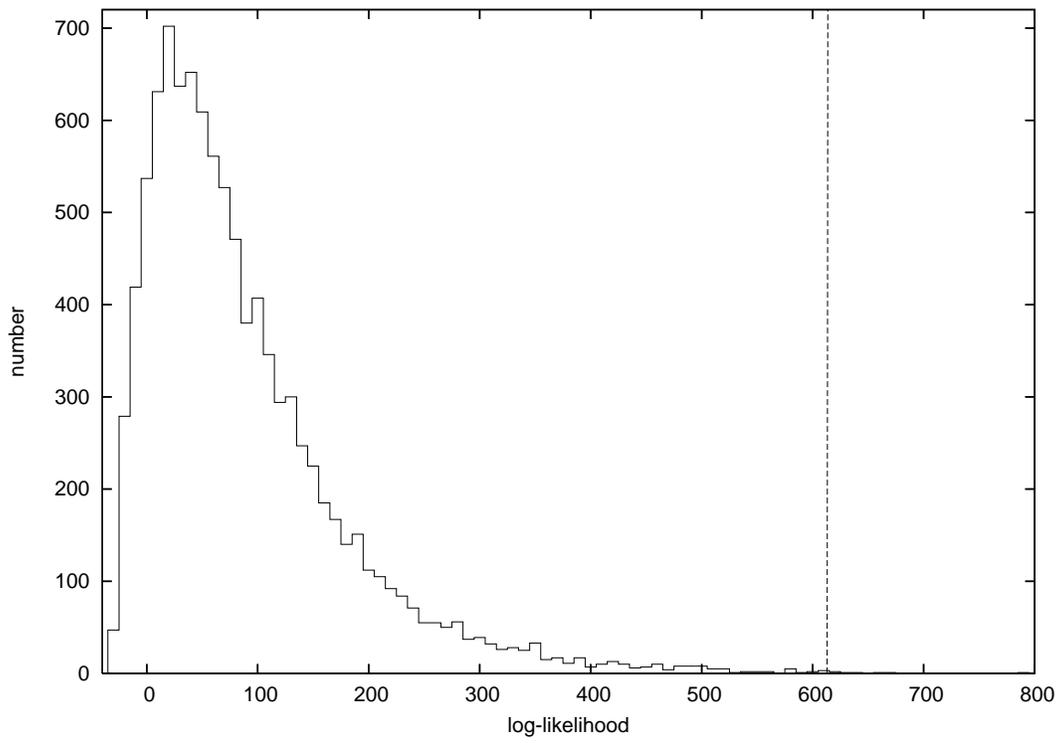}
\caption{The log-likelihood distribution coming from the $10^5$ synthetic datasets 
generated from one single value of $p$ (solid line); 
the dashed vertical line indicates the log-likelihood for the real data. In $99.92\,\%$ of the cases 
the log-likelihood of the synthetic data is smaller than the one coming from our measured sample.} 
\label{fig:loglikelihood}
\end{center}
\end{figure}

To test our null-hypothesis that the observed distribution of $p$ can be obtained from a parent distribution with a single central value of $p$, 
we generate $10^5$ different synthetic datasets for $p$ for the 10 bursts in our sample by taking random numbers 
from probability distributions that are described by the most likely value of $p=2.04$ and the $1\,\sigma$ uncertainties 
in the 10 values of $p$. We then take 10 values of $p$ coming from these synthetic datasets (one random number for each measurement), 
calculate the most likely value of $p$ and the accompanying log-likelihood. 
The resulting distribution of log-likelihood values is plotted in Fig. \ref{fig:loglikelihood}, 
together with the minimal log-likelihood of $613.6$ coming from the measurements. 
We find that in $99.92\,\%$ of the cases the log-likelihood of the synthetic data is smaller than the one 
coming from the measured $p$-values. \emph{This leads to the conclusion that the hypothesis, that the distribution of $p$ from our sample 
can be obtained from a parent distribution with a single central value of $p$, is rejected at the $3\,\sigma$-level.} 

This result challenges the theoretical studies of particle acceleration which claim that there is one universal value of $p$ 
\citep[e.g.][]{kirk2000:apj542:235,achterberg2001MNRAS32:393}, and is consistent with
similar findings by \cite{Shen2006MNRAS371} from fits to the prompt emission of a
sample of Burst And Transient Source Experiment ({\it BATSE}) GRBs.
\citet{barringNucPhysB2004:136:198} shows that this often quoted value of $p=2.23$ is a special case, 
with particular assumptions on hydrodynamic quantities, like the compression ratio of the upstream and downstream velocities, 
on the influence of the magnetic field on the dynamics, and on electron scattering angles. 
He claims that there is a large range of possible values for $p$ of $1.5-4$. 
We derive constraints on the width of the distribution of $p$-values below.
\clearpage
\clearpage
From Table \ref{table:finalresults} and Fig. \ref{pvalues} it can be seen
that the log-likelihood is dominated by the value for $p$ of GRB\,990123, and after that by GRB\,980519. 
In fact, just the two $p$-values of GRB\,990123 and GRB\,980519 are different by $12\,\sigma$, 
and thus they alone refute the constant $p$ hypothesis. 
We also examined the strength of the evidence against a universal $p$ from samples excluding one of the 10 bursts in our sample. 
Only in cases where either GRB\,990123 or GRB\,980519 are excluded is the significance less than $3\,\sigma$, 
but even then the hypothesis can be rejected at the $\sim 2\,\sigma$-level. 

Since we now know that the measured values of $p$ from our sample are not coming from a parent distribution with a single central value of $p$, 
we can put constraints on the width of the parent distribution of $p$-values. 
We do this by introducing an intrinsic scatter on the most likely value of $p=2.04$: 
we generate $10^5$ synthetic datasets by first taking random numbers from normal distributions with a central value of $p=2.04$ 
and a certain scattering $\sigma_{\mathrm{scat}}$; we then take random numbers from the probability distributions 
that are centered at these scattered $p$-values and have a width given by the measurement uncertainties. 
Performing the same log-likelihood analysis as described above for varying $\sigma_{\mathrm{scat}}$, 
we find that the $3\,\sigma$ result holds for $\sigma_{\mathrm{scat}}>0.03$; at the $1\,\sigma$ level this is $\sigma_{\mathrm{scat}}>0.24$. 
We can also constrain the upper limit on the scatter in the parent distribution of $p$-values from our sample. 
This can be derived by determining for which $\sigma_{\mathrm{scat}}$ the log-likelihood of the measured $p$-values 
is smaller than the one from the synthetic dataset analysis in $99.73\,\%$ of the $10^5$ cases. 
This upper limit for $\sigma_{\mathrm{scat}}$ is $1.40$ at the $3\,\sigma$ level; at the $1\,\sigma$ level this is $\sigma_{\mathrm{scat}}<0.45$. 
In summary, from our sample we constrain the intrinsic width of the parent distribution of $p$-values to be $0.24<\sigma_{\mathrm{scat}}<0.45$ 
at the $1\,\sigma$ level and $0.03<\sigma_{\mathrm{scat}}<1.40$ at the $3\,\sigma$ level

Values of $p=2.2$ have been widely quoted as a typical
number and the deviations from this interpreted as due to the external
environment or further energy injections from the source \citep[e.g.][]{berger2003:apj588}.
\citet{meszaros1998:apj499:301} showed that for a single value of \p,
variations in the jet energy per solid angle, i.e. structured jets, could lead to a range of
lightcurve decays; which can also be produced if a set of intrinsically similar, 
structured jets are viewed at various off-axis angles \citep{rossi2002:mnras332:945}. 
The study of such structure in the jet and viewing angle dependence is, however, beyond the scope of this work.

\citet{Johannesson2006:apj640:L5} claim that from broken power law fits
the value of $p$ is overestimated from the pre-break lightcurve slope while
being underestimated from the post-break slope, if the temporal coverage is limited.
The results presented in that paper were only for a homogeneous ambient medium;
they claim that in a wind medium the $p$ value from the pre-break lightcurve slope
does not show a systematic deviation.

More recently, breaks in lightcurves, both optical and X-ray, which
would generally have been attributed to jet-breaks, have been found to
be chromatic, in sharp contrast to the picture of a jet-break which should be achromatic. As such,
several previously claimed jet-breaks may have to be revised;
unfortunately, there is no relevant data to confirm or reject these
claimed jet-breaks, since these have been found only in one waveband
(mostly the optical, owing to the lack of dense monitoring in X-rays in the
pre-{\it Swift} era). As such,
the mention of jet-breaks in this paper has been taken at face value,
but with the caveat mentioned here.

If we compare the values for \p\ calculated here with those calculated from
the X-ray spectra alone \citep{stratta2004:apj608:846}, we find they agree
within the 90\% errors except for the bursts GRB\,970508 and GRB\,990510. 
With this method we reduce the average 90\% error on $p$ from $\pm$0.58 
\citep{stratta2004:apj608:846} to $^{+0.12}_{-0.20}$, 
and the values obtained are also likely to be more robust given that consistency between the nIR, optical, UV and X-ray is required.

\cite{Shen2006MNRAS371} have measured the minimum possible width of the electron energy
index distribution for X-ray afterglows of {\it BeppoSAX} bursts taken from
\cite{DePasquale2006AandA455},
by fitting single absorbed power laws to the spectra and
deriving \p. They found that the narrowest possible distribution is consistent
with a delta function within the 1$\sigma$ errors, in contradiction to their
findings from a {\it Swift} sample of X-ray afterglows and from the prompt emission
of a {\it BATSE} GRB sample. They comment that the relatively larger errors on the
X-ray slopes of the {\it BeppoSAX} sample compared with the {\it Swift} sample are likely
allowing for an apparently narrower distribution. 
They calculate that the distribution of $p$ for the {\it Swift} sample of X-ray
afterglows \citep{obrien2006:apj647:1213} has a width of $0.34\pm 0.07$.
We note that \cite{Shen2006MNRAS371} have derived distributions of $p$ from X-ray afterglows only,
which makes it impossible to know whether the cooling frequency lies above or below the X-ray regime,
and can only be determined if there are accompanying optical measurements.

\citet{panaitescu2002:apj571} calculated jet parameters for a sample of 10 GRBs
including several studied here, using broadband observations including radio data
when possible. In their model, based also on the fireball model,
they assume uniform jets (no structure across the jet) and the the energy
parameters $\epsilon_e$ and $\epsilon_B$ are constant, and finally they assume
the observer is located on the jet symmetry axis. Our spectrally-derived
values are consistent with theirs at the 3$\sigma$ level or better for GRBs 970508,
980519, 990510 and 000926. There is no agreement for GRB\,990123. 
They found a spread in $p$ values as do we, but with half the values lying below 2, and a
mean value of $p$ = 1.92$^{+0.28}_{-0.26}$ (2$\sigma$).

\citet{chevalier2000:ApJ536:195} carried out a similar study of broadband afterglow
data, and their estimates for $p$ are in agreement with ours for the GRBs
970228, 970508, 980519 and 990510, and disagree for GRB\,990123 (they do not quote
errors per source but estimate errors to be $\sim$0.1). They conclude that their sample shows a range in the values of
$p$ which is not consistent with a single value.

We note that in all these studies the results are dependent upon the chosen
model and simplifying assumptions.


\subsection{The Circumburst Medium}

The profile of the circumburst medium is a particularly important parameter in studying the progenitors 
of GRBs. In the case of long-soft bursts the progenitor is a massive star that is expected 
to have had a massive stellar wind in earlier phases of its evolution, and one would expect to see 
a signature of that wind in the afterglow lightcurves. Evidence for a stellar wind in the form of fast 
outflowing absorption lines within restframe UV spectra has been seen in a small number of cases, 
the best example being GRB\,021004 \citep[e.g.][]{schaefer2003:ApJ588:387,starling2005:MNRAS360:305}. 
This does not mean, however, that a density profile with $n\propto r^{-2}$ is always expected, since this assumes 
a constant mass-loss rate and a constant wind velocity. Changes in mass-loss rate and also interactions 
of the wind with the interstellar medium can alter the ambient medium profile 
\citep[e.g.][]{ramirez-ruiz2005:ApJ631:435,vanmarle2006:a&a460:105}. 

In the first broadband modeling attempts the ambient medium was assumed to be the homogeneous interstellar 
medium, which was consistent with the derived particle densities. However, since the progenitors of at least 
a fraction of all GRBs are now known to be massive stars and the blastwave is situated at $\sim 10^{17}$ cm 
at $\sim 1$ day after the burst, a massive stellar wind profile is expected. Nonetheless, the medium that the 
blastwave is probing could still be homogeneous, because of the emergence of a reverse shock in the wind when 
the wind meets the interstellar medium \citep[see e.g.][]{wijers2001:grbaconf306,ramirez-ruiz2005:ApJ631:435}. 
This shocked wind turns out to become homogeneous and, for typical physical parameters derived from afterglow modeling, 
the blastwave encounters the transition from the stellar wind to this homogeneous shocked medium at $\sim 1$ day 
\citep[see e.g.][]{pe'er2006:ApJ643:1036}. The actual time of the transition, which would be detectable in the afterglow 
lightcurves, depends for instance on the mass-loss rate and the density of the interstellar medium, which are 
both not really well constrained for most GRBs. In our sample we do not see this kind of transition in the 
optical lightcurves in which there is a break. It has been claimed for some bursts, 
for instance GRB\,030329 \citep{pe'er2006:ApJ643:1036}, that this transition is observed, 
but the often complex structure of the lightcurves confuses the determination of such a transition. 

Another way to obtain a constant density from a massive stellar wind is in
the region after the wind termination shock. The distance to the termination
shock can be very large, but recent observations of two Wolf-Rayet binaries
has suggested that this distance could be several times smaller if the wind is
asymmetric. \citet{eldridge2007:astroph1707} shows that wind asymmetry probably exists in two
systems, that can be caused for example by rotation, which is expected for GRB
progenitors in the framework of the collapsar model in order to retain enough angular momentum. 
If the asymmetry exists for the entire stellar lifetime, then a closer termination shock and asymmetric supernova 
may be expected, increasing the chances of observing an afterglow traversing a constant density medium.

In our sample of 10 GRBs there are four sources that are consistent with an $r^{-2}$ wind medium, with 
relatively small uncertainties, namely GRBs 971214, 990123, 000926, and 010222. There are three 
GRBs which are not consistent at a $90\,\%$ confidence level with a wind
medium, GRBs 970508, 980519, and 990510, although for GRB\,970508 caution is warranted with the interpretation of the lightcurves; 
and for the other four bursts the uncertainties are too large to constrain the ambient medium profile. 
We contrast our findings with \citet{panaitescu2002:apj571}, who, in broadband fits to the data of 10 bursts, 
found that a wind-like medium was preferred over a homogeneous medium in only one case: GRB\,970508. For
this particular burst our analysis provides a value of $p$ which is an equally good fit to wind 
or uniform medium predictions for $\nu<\nu_c$ from the spectra and lightcurves, but the closure relations are obeyed (at the 2$\sigma$ level for both cases) if $\nu>\nu_c$. They find circumburst densities of order 0.1--100 cm$^{-3}$ 
for most sources, which they argue demonstrates the surrounding medium does not have, in general, the $r^{-2}$ profile expected for the unperturbed wind of a massive GRB progenitor.

The association of long-soft GRBs with Ib/c supernovae was first suggested for
GRB\,980425\,/\,SN\,1998bw by Galama et al. (\citeyear{galama1998:nature395:670}), and confirmed by the discovery of 
GRB\,030329\,/\,SN\,2003dh \citep[e.g.][]{hjorth2003:nature423:847}. Therefore, it is useful to compare the 
circumburst medium characteristics derived from GRB afterglows and from radio observations of 
supernovae, which also trace the density profile of the surroundings of these massive stars. 
Around radio supernovae $r^{-2}$ density profiles have been found, but also in some cases significantly 
flatter behavior of $\sim r^{-1.5}$ in SN\,1993J and SN\,1979C \citep[for a review
on radio supernovae see][and references therein for individual supernovae]{weiler2004:NewAR48:1377}. 
In the latter case a transition from $r^{-2}$ to $r^{-1.4}$ was even observed. This flatter density profile 
can be attributed to changes in the mass-loss rate of the massive star in some phases of its evolution. 
The three sources in our sample of GRBs with a density profile flatter than $r^{-2}$ are possible examples 
of the relativistic blastwave ploughing its way through a region of the circumburst medium which is 
affected by changing mass-loss rates. So although in Table \ref{table:finalresults} we have described them 
as GRBs with a homogeneous ambient medium, this is not necessarily the
case. Especially for GRBs 970508 and 980519 this is a possibility, but the
uncertainties on $k$ are too large to distinguish a homogeneous from a flattened wind medium. GRB\,990510 has smaller uncertainties and seems to be more consistent 
with a homogeneous medium, especially since the upper limit on $k$ is $\sim 1.0$, which is much flatter than 
what is observed in radio supernovae.

The {\it Swift} satellite now provides us with substantially greater coverage
of a large number of X-ray afterglows ($\sim$100 per year) and often with high
quality data from which to measure the spectral and temporal slopes. However,
few of these also have substantial optical follow-up. The combination of X-ray
and optical data helps determine the position of the cooling break and obtain
accurate spectral slopes which provide the value of \p. For the derivation of
$k$ in this study, we have found the optical temporal data most constraining. For this
reason, and for the confirmation of achromatic jet breaks it is essential that
such late-time optical data be obtained for as many well sampled {\it Swift} X-ray afterglows
as possible.


\section{Conclusions}

We have measured the injected electron energy distribution index, \p, in the
framework of the blastwave model with some assumptions which include
on-axis viewing and standard jet structure, constant blastwave energy and no
evolution of the microphysical parameters. We have also measured the density profile of the circumburst medium, $n(r) \propto r^{-k}$, 
from simultaneous spectral fits to the X-ray, optical and nIR afterglow data
of 10 {\it BeppoSAX} GRB afterglows. 

A statistical analysis demonstrates that the distribution of $p$ values in this sample 
is inconsistent with a single value for $p$ at the $3\,\sigma$ certainty, 
which is at variance with many theoretical studies of relativistic particle acceleration. 
We constrain the width of the parent distribution of $p$ values and find it to be of the order of a few tenths, 
with $p=2.04^{+0.02}_{-0.03}$ as the most likely $p$-value in our sample.

We measure the distribution of the local density parameter $k$, generally only
assumed to be 0 or 2, and we find that the majority of GRBs for which we can
constrain $k$ well are consistent with a wind-like circumburst medium. One source (GRB\,990510) is
clearly, i.e. $>3\,\sigma$ certainty, inconsistent with this picture and fits instead a homogeneous medium. 
These results are consistent with the expectations of at least a subset of GRBs 
originating from massive stars, which have a lot of mass-loss in the form of a surrounding stellar wind. 
We have discussed the possibility of values of $k<2$ within the stellar wind framework. 

The method presented here provides a way to study the distribution of blastwave parameters
for a sample of GRBs, and allows estimates to be derived when insufficient data are available 
for a full time-dependent solution. 
In the current {\it Swift} era the method is equally well applicable, although one would have to 
ensure that the data are in the afterglow domain, i.e. not contaminated by prompt emission 
or late-time energy-injection. A decent sampling of the optical SED and lightcurve, 
more difficult with the average fainter {\it Swift} afterglow sample, is crucial to constrain 
the temporal slopes and cooling break frequency, which in turn are necessary to determine $p$ and $k$.


\acknowledgments

We thank Dipankar Bhattacharya for useful discussions, and the referee
and Alexander Kann for useful comments on the manuscript. 
We thank Mike Nowak for his assistance with {\small ISIS}, Nanda Rea
for assistance with the {\it BeppoSAX} data reduction and Martin Heemskerk for
his help with the running of both {\small ISIS} and {\small SAXDAS}. 
This research has made use of {\small SAXDAS} linearized and cleaned event
files produced at the {\it BeppoSAX} Science Data Center.
The authors acknowledge benefits from collaboration within the Research
Training Network `Gamma-Ray Bursts: An Enigma and a Tool', funded by the EU
under contract HPRN-CT-2002-00294. RLCS and ER acknowledge support from PPARC. 
RAMJW, PAC and KW thank NWO for support under grant 639.043.302.

\bibliographystyle{apj}
\bibliography{references}


\end{document}